\documentclass[conference]{IEEEtran}
\usepackage{graphicx}
\usepackage{amsfonts}
\usepackage{amsmath}
\usepackage{amssymb}
\usepackage{cite}
\usepackage{algorithm}         
\usepackage{algorithmic}         
\usepackage{verbatim}
\usepackage{commands11}

\usepackage{lipsum}

\newcommand\blfootnote[1]{%
  \begingroup
  \renewcommand\thefootnote{}\footnote{#1}%
  \addtocounter{footnote}{-1}%
  \endgroup
}

\graphicspath{{fig/}}

\renewcommand{\B}[1]{\boldsymbol{#1}}

\begin{document}
%

\title{Distributed Ranging and Localization for Wireless Networks via Compressed Sensing}

\author{\IEEEauthorblockN{Ming Gan\IEEEauthorrefmark{1},
Dongning Guo\IEEEauthorrefmark{2} and Xuchu Dai\IEEEauthorrefmark{1}}

\IEEEauthorblockA{\IEEEauthorrefmark{1}Department of Electronic Engineering and Information Science, \\
University of Science and Technology of China, Hefei, Anhui, China}
\IEEEauthorblockA{\IEEEauthorrefmark{2} Department of Electrical Engineering and Computer Science, \\
Northwestern University, Evanston, IL, USA} }

\maketitle

\begin{abstract}
  Location-based services in a wireless network require nodes to know
  their locations accurately.
  Conventional solutions rely on contention-based medium access,
  where only one node can successfully transmit at any time in any
  neighborhood.
  In this paper, a novel, complete, distributed ranging and
  localization solution is proposed, which let {\em all} nodes in the
  network broadcast their location estimates and measure distances to
  {\em all} neighbors simultaneously.
  An on-off signaling is designed to overcome the physical half-duplex
  constraint.
  In each iteration,
  all nodes transmit simultaneously, each broadcasting
  codewords describing the current location estimate.  From
  the superposed signals from all neighbors, each node
  decodes their neighbors' locations and also estimates
  their distances using the signal strengths.  The
  node then broadcasts its improved location estimates in the
  subsequent iteration.
  Simulations demonstrate accurate localization throughout a large
  network over a few thousand symbol intervals, suggesting much higher
  efficiency than conventional schemes based on ALOHA or CSMA.\blfootnote{This work was supported in part by the NSF under Grant ECCS-1231828}
\end{abstract}

\section{Introduction} 
\label{s:intro}
Many location-based wireless services have emerged over the last
decade, e.g., emergency services, tracking services, proximity
advertising, and smart home monitoring~\cite{steiniger2006foundations,
  mohapatra2008survey, surie2005wireless}.
Such applications require wireless nodes to obtain their own locations
accurately, which is referred to as {\em localization}.  Localization
is also crucial in many mobile ad hoc networks.
The Global Positioning System (GPS) provides a straightforward
solution for localization.
However, GPS may be unavailable, e.g., indoor or in between high
rises, too costly or not accurate enough in many situations.
In this paper, we study the problem of distributed localization by
letting nodes transmit signals to and receive signals from its one-hop
neighboring nodes, referred to as its {\em neighbors}.

For ease of discussion, the wireless network is assumed to be
connected, and has at least three {\em anchors}, namely, nodes who
know their own exact locations.
Nodes who do not know their own locations to begin with are referred
to as \emph{clients}.
In the absence of anchors, the techniques proposed in this paper still apply,
whereas the nodes can only be located subject to rigid translation,
rotation and reflection.
In general, a node determines its own location using location and
distance information from its neighbors.
The distances can be estimated based on the
time of flight, signal strength or other
measurements~\cite{kim2009performance, hwang2011AVSR, fu2007ranging}.


Given the neighbors' location estimates and distances, each node can solve an
optimization problem to estimate its own location~\cite{srirangarajan2008distributed,qingjiang2010distributed,chiu2012robust}.
Although the clients and their neighbors (except for the anchors)
generally do not know their true locations at the beginning, by
iteratively improving their estimates and updating their neighbors with
their new location estimates, eventually all connected clients in the
network are expected to converge to good estimates of their locations.

To accomplish distributed ranging and localization requires two
lower-level functions:
i) Each node should be able to communicate to
its neighbors about its own location or its location estimate;
ii) Each node should be able to measure the (approximate) distances to
its neighbors.

Conventional solutions for exchanging message and ranging
are based on contention-based random access, such as ALOHA or
carrier-sensing multiple access (CSMA).
Localization is usually solved in the network layer or above.
In particular, each node can only receive useful signals from one neighbor at a time
for ranging or exchanging location information.  
Multiple transmissions in a neighborhood collide
and destroy each other.
Hence successful localization requires many retransmissions over many frame
intervals.

In this work, we point out that ranging
and localization are fundamentally physical-layer issues
and propose novel physical-layer techniques for solving them.
In particular, each node can exchange messages with multiple neighbors
and can measure its distances to multiple neighbors in a single frame
all at the same time, as long as transmissions from different nodes
are distinguishable.
In lieu of colliding with each other, simultaneous transmissions
superpose at the receiver.
As a toy example, consider three nodes who are neighbors of each other.
In a given frame, node 2 and node 3 transmit signals
$\B{S}_2$ and $\B{S}_3$, respectively.
The received signal of node 1 during this frame is $a_{12}
\B{S}_2 + a_{13}\B{S}_3$, where $a_{12}$ and $a_{13}$ denote
the corresponding received signal amplitudes.
If the signals $\B{S}_2$ and $\B{S}_3$ are sufficiently different in
some sense,
node~1 can estimate the amplitudes $a_{12}$ and $a_{13}$
at the same time, and can in turn infer about the corresponding
distances to nodes 2 and 3.
Further, if  $\B{S}_2$ and $\B{S}_3$ are codewords that bear
information about the location of node 2 and node 3, respectively,
node 1 can also decode their locations at the same time.
Zhang and Guo~\cite{zhang2013virtual} first pointed out that the problem
of exchanging messages in a wireless network is fundamentally a
problem of {\em compressed sensing} (or {\em sparse recovery}).
Compressed sensing~\cite{donoho2006compressed, candes2008introduction,
 candes2005decoding} studies the problem of efficiently recovering
a sparse signal based on relatively few measurements made through a
linear system.
In this work, we carry out location information exchange and ranging
jointly using compressed sensing techniques.

%

One challenge in wireless networks is the half-duplex constraint,
where a node cannot transmit and receive useful signals at the same time
over the same frequency.  To resolve this, we employ the
rapid on-off division duplex (RODD) signaling proposed
in~\cite{zhang2013virtual, guo2010virtual, zhang2013neighbor}.
The key idea is to let each node randomly divide a frame (typically of a few
hundred to a few thousand symbols) into on-slots and off-slots, where the node
transmits during the on-slots and listens to the channel during the
off-slots.
A node receives a superposition of neighbors' transmissions through
its off-slots. The received messages can be decoded as long as the
frame is sufficiently long.

The proposed physical-layer technique assumes that node transmission
are synchronized. 
Local-area synchronization can be achieved using a common source of
timing, such as a beacon signal, or using a distributed consensus
algorithm~\cite{schizas2008consensus,simeone2008distributed}.
To maintain synchronization requires an upfront cost in the operation
of a wireless network. The benefit, however, is not limited to the
ease of ranging and localization, but includes improved efficiency in many
other network functions.  More discussions on the synchronization issue are found
in~\cite{zhang2013virtual,guo2010virtual}.


The proposed ranging and localization algorithm is validated through
simulations.  A network of one hundred Poisson distributed nodes is
considered, where transmissions are subject to path loss and additive
Gaussian noise.  Nodes also interfere with each other.  The iterative
algorithm is carried out under rather realistic assumptions.
Numerical results show that about 10 iterations suffice, where each
iteration requires 1200 symbol transmissions.  The total number of
symbol transmission is about 12,000 symbols.  (As a reference, one
WiFi frame consists of several thousand symbols.)
Thus the proposed scheme is much more efficient than random access
schemes, where many more transmissions and retransmissions are needed
due to collisions.

The remainder of the paper is organized as follows:
Section~\ref{s:model} describes the channel and network models.  The
distributed ranging and localization algorithm is described in
Section~\ref{s:loc}.  Section~\ref{s:num} presents numerical results
and Section~\ref{s:con} concludes the paper.

\section{Channel and Network Models}
\label{s:model}

\subsection{Notation}
In this paper, upper case letters denote random variables and the
corresponding lower case letters denote their realizations or
estimates.
Vectors and matrices are denoted by bold-face letters.
$\expect{\cdot}$
denotes the expectation over the random variables within the bracket.
$\reals$ denotes the real number set.
$|\cdot|$ denotes the absolute value of a number
or
the cardinality of a set.
$\|\cdot\|$ represents the Euclidean norm of a vector.

\subsection{Linear Channel Model}

Let $\Phi=\{Z_i\}_i$ denote the set of nodes on the two-dimensional
plane.
We refer to a node by its (random) location $Z_i \in \reals^2$.
Suppose all transmissions are over the
same frequency band. Let time be slotted and all nodes be
perfectly synchronized over each frame of $M_s$ slots. For simplicity,
each slot consists of one symbol.
Suppose node $Z_i$ wishes to broadcast an  $l$-bit message $\omega_i$
to its neighbors.
Let the signature $\B{S}_i\left(\omega_i\right)$ denote the
$M_s$-symbol codeword transmitted by node $Z_i$ over symbol intervals
1 through $M_s$, whose entries take the values in $\{-1,0,1\}$. During
slot $m$, the node $Z_i$
listens to the channel and emits no energy if the $m$-th bit of the
codeword $\B{S}^m_i =0$; otherwise, the node $Z_i$ transmits a
symbol. The design of the on-off signatures
will be discussed in Section~\ref{s:loc}.

The physical link between any
pair of nodes is modeled as a fading channel. Let the path loss satisfy a power law with exponent
$\alpha$. Without loss of generality, we focus on node $Z_0$.
The signal received by node $Z_0$, if it could listen over the entire
frame, can be expressed as
\begin{align}\label{1}
\widetilde{\B{Y}}_0 = \mathop{\sum}_{Z_i \in \Phi \backslash\{Z_0\}} \sqrt{\gamma}\, h_{0i} R^{-\alpha/2}_{0i} \B{S}_i\left(\omega_i\right) + \widetilde{\B{W}}_0,
\end{align}
where $\Phi \backslash\{Z_0\}$ denotes the set of all nodes except $Z_0$,
$\gamma$ denotes the nominal signal-to-noise ratio (SNR), $R_{0i}=\|Z_0-Z_i\|$ is the distance between the nodes $Z_0$ and
$Z_i$,  $h_{0i}$ denotes the small-scale fading coefficient,
and $\widetilde{\B{W}}_0$ is additive white noise vector whose components
follow the unit circularly
symmetric complex Gaussian distribution $\mathcal{CN} \left(0, 1\right)$.
For any pair of nodes, e.g., $Z_i$ and $Z_j$, the channel gain between
them can be expressed as  $|h_{ij}|^2 R^{-\alpha}_{ij}$.  Channel
reciprocity is a given, i.e., $h_{ij}=h_{ji}$.

Let us denote the set of neighbors of node $Z_j$ as
\begin{align}  \label{eq:NZi}
  \mathcal{N}\left(Z_j\right) = \left\{Z_i\in \Phi: |h_{ij}|^2
    R^{-\alpha}_{ij} \geq \theta, i\neq j\right\}
\end{align}
where $\theta$ is the threshold. The reasons for not defining the neighborhood purely based on the geometrical closeness as
~\cite{srirangarajan2008distributed,qingjiang2010distributed} are as
follows:
i) The channel gain plays
a key role in determining the connectivity between a pair
of nodes; ii) The attenuation generated by path loss and fading can
not be separated within one frame.

Transmissions for non-neighbor are accounted for as part of the additive Gaussian noise. Hence, \eqref{1} can be rewritten as
\begin{align}\label{eq:y0}
  \widetilde{\B{Y}}_0
  =
  \mathop{\sum}_{Z_i \in \mathcal{N}(Z_0)}
  \sqrt{\gamma} \, U_{0i}  \B{S}_i\left(\omega_i\right) + \overline{\B{W}}_0,
\end{align}
where the channel coefficient
\begin{align}  \label{eq:U0i}
  U_{0i}=h_{0i} R^{-\alpha/2}_{0i}
\end{align}
and the components of $ \overline{\B{W}}_0$ follow  $\mathcal{CN} \left(0, \sigma^2\right)$.
The variance $\sigma^2$ results from the additive noise
$\widetilde{\B{W}}_0$ as well as the aggregate interference caused by
non-neighbors, which will be given in Section~\ref{s:enc} because it
depends on the transmission scheme.

\subsection{Node Distribution and Neighborhoods} 

Suppose all nodes are distributed across the two-dimensional plane according to a homogeneous Poisson point process with intensity $\lambda$, which is
the frequently used to study wireless network (see~\cite{baccelli2009stochastic} and references therein). The number of nodes in
any region of area $A$ is a Poisson random variable with mean $\lambda
A$.

We next derive the relationship between the average number of
neighbors a node has and the threshold $\theta$ that defines the
neighborhood.  These parameters shall be used by the receiver to be
described in Section~\ref{s:loc}.
Without loss of generality, we drop the indices of the pair of nodes
of interest.
By definition of a neighbor, the channel gain  must satisfy $|h|^2 R^{-\alpha}\geq \theta$, i.e., $R \leq \left(|h|^2/\theta\right)^{1/\alpha}$.
Under the assumption that all nodes form a Poisson point process, for given $|h|^2$, this arbitrary neighbor node $Z_i$ is uniformly distributed
in a disc centered at node $Z_0$ with radius $\left(|h|^2/\theta\right)^{1/\alpha}$. Hence the conditional distribution of $R$ given $|h|^2$ is
given by
\begin{align}\label{3}
P\left(R \leq r \big| |h|^2\right)=\min\left\{1, r^2\left(\frac{\theta}{|h|^2}\right)^{\frac{2}{\alpha}}\right\}.
\end{align}
Now for every $u \geq \sqrt{\theta}$, using~\eqref{3}, we have
\begin{align}\label{4}
  P(|h|^2 R^{-\alpha} \geq u^2 )
  &= \expsub{|h|^2} {P\left(R \leq \left(\frac{|h|^2}{u^2}\right)^\frac{1}{\alpha} \right)\bigg| |h|^2 } \nonumber
\\
&= \expsub{|h|^2} { \left(\frac{|h|^2}{u^2}\right)^{\frac{2}{\alpha}}  \left(\frac{\theta}{|h|^2}\right)^{\frac{2}{\alpha}} } \nonumber
\\&= \frac{\theta^{\frac{2}{\alpha}}}{u^{\frac{4}{\alpha}}}.
\end{align}
Therefore, for an arbitrary neighbor $Z_i$ of node $Z_0$,
the probability density function (pdf) of $|U_{0i}|$, which is the
amplitude of the channel coefficient, is
\begin{align}\label{eq:p(u)}
p(u)=\left\{
\begin{aligned}
&\frac{4}{\alpha}\frac{\theta^{\frac{2}{\alpha}}}{u^{\frac{4}{\alpha}+1}}, &u\geq \sqrt{\theta} ;\\
&0, &\text{otherwise.}
\end{aligned}
\right.
\end{align}

In fact, fading coefficient vector $\mathcal{G}_i=\left\{|h_{ij}|^2\right\}_j$
for all $j\neq i$ can be regarded as an independent mark of node $Z_i$, so that
$\tilde{\Phi}=\left\{\left(Z_i, \mathcal{G}_i\right)\right\}_i$ is a
marked Poisson point process. Let
$\hat{\Phi}=\left\{\left(Z_i,\mathcal{G}_i\right)\right\}_i\backslash\left\{\left(Z_0,\mathcal{G}_0\right)\right\}$ denote the pair set $\tilde{\Phi}$
excluding the pair $\left(Z_0,\mathcal{G}_0\right)$.  By the
Campbell's theorem~\cite{baccelli2009stochastic},
$\hat{\Phi}$ is also marked Poisson
point process with intensity $\lambda$, the average number of
neighbors of node $Z_0$ can be obtained using~\eqref{eq:p(u)} as
\begin{align}\label{6}
  c
  &= \expsub{\hat{\Phi}} { \mathop{\sum}_{\left(Z_i, \mathcal{G}_i\right)\in \hat{\Phi}} \mathbf{1}\left( |h_{0i}|^2 R^{-\alpha}_{0i} \geq \theta \right) } \nonumber
\\
&=2\pi\lambda \int^\infty_0 \int^\infty_0  \mathbf{1}\left( \bar{h} r^{-\alpha} \geq \theta \right) r e^{-\bar{h}} drd\bar{h} \nonumber
\\
&=\frac{2}{\alpha}\pi\lambda \theta^{-\frac{2}{\alpha}}\Gamma\left(\frac{2}{\alpha}\right),
\end{align}
where $\mathbf{1}(\cdot)$ and $ \Gamma(\cdot)$ are the indicator function and Gamma function, respectively.

\section{The Localization Problem}
\label{s:loc}

It is assumed that every node in the network has already discovered
its set of neighbors defined by~\eqref{eq:NZi}, e.g., by using
techniques proposed in~\cite{zhang2013virtual} or~\cite{borbash2007asynchronous}.
That is to say, the node has acquired each neighbor's network
interface address (NIA).
It is fair to assume that the node also knows the codebook used by
each neighbor for transmission.
This is easily accomplished, e.g., by letting each node generate its
codebook using a pseudo random number generator with its NIA as the
seed.

The network $\Phi$ consists of a subset $\Phi_a$
of anchors,
who know their own locations, and a subset $\Phi_c$
of
clients, who wish to estimate their locations.
%
%
The problem of localization is to let all nodes in the network
estimate their respective locations through some
transmissions between neighboring nodes.
We are only interested in clients who are connected to the dominant
component of the network.


\section{The Proposed Distributed Algorithm}
\label{s:algo}

In this section, we describe an iterative, distributed ranging and
localization algorithm carried out by all nodes in the network.  As
discussed in Section~\ref{s:intro}, node transmissions are
synchronized in each local area.

The algorithm is roughly describes as follows: Each iteration
corresponds to two frame intervals, during which a node may transmit
two codewords to describe the two coordinates of its location,
respectively.
During the same two frame intervals, the node also receives the
superposition of its neighbors' transmissions.  The node then
decodes
the location information from its neighbors as well as infers the
distances to the neighbors based on the received signal strengths.
Using the locations of and distances to the neighbors, the node can
estimate its own location, which it then broadcasts to its neighbors in the
subsequent iteration.
Note that the anchors always broadcast their actual locations.
Although the clients do not know their own locations initially, it is
expected that nodes near the anchors will first locate themselves with
some accuracy, which, in turn, help their neighbors to locate themselves.
After sufficient number of iterations, all clients in the network shall
have a good estimate of their locations, subject to uncertainty
due to noise, interference, and node connectivity.

In the following, we first discuss each step of the localization algorithm
in detail, and then summarize the overall procedure using a flow chart
(Fig.~\ref{f:chart}).



\subsection{Quantization of Location}

Suppose the network is confined to the area $[0,A]\times[0,A]$ on the
plane.  Although the physical location of a node is described by two
real numbers on $[0,A]$, the node can only communicate with its
neighbors with finite precision due to noise and interference.  Let
the node quantize its location $z$ to $(\omega,\nu)$, where $\omega$
and $\nu$
are $l$-bit strings.  The quantization step size is equal to
$\Delta=2^{-l}A$.
It suffices to quantize to precision finer than the expected
localization error.


\subsection{Encoding}
\label{s:enc}


Let the two quantized coordinates of node $Z_i$'s location, denoted by
$(\omega_i,\nu_i)$ be sent over two separate frames.
Following~\cite{zhang2013virtual}, we let node $Z_i$ randomly and
independently generate a codebook of $2^l$ codewords, each consisting of
$M_s$ symbols.  Specifically, let the $2^l M_s$ entries of the codebook be
independent and identically distributed, where each entry is equal to
0 with probability $1-q$ and is equal to 1 and $-1$ with probability
$q/2$ each.
To broadcast coordinates $(\omega_i,\nu_i)$ to its neighbors, node
$Z_i$ transmits over the first frame interval the $\omega_i$-th
codeword from its codebook, denoted by $\B{S}_i(\omega_i)$, and
transmits $\B{S}_i(\nu_i)$ over the second frame interval.

\subsection{The Received Signals}

Without loss of generality, we focus on the signals received by node
$Z_0$ and assume its neighbors are indexed by $1,\dots,K$, where
$K=|\mathcal{N}(Z_0)|$.
The total number of signatures owned by all the neighbors is $N=2^l
K$.
Node $Z_0$ basically receives a superposition of $K$ codewords, one
chosen from each codebook.  What complicates the matter somewhat is
the fact that $Z_0$ can only listen to the channel through its own
off-slots due to the half duplex constraint.

The number of $Z_0$'s off-slots, denoted by $M$, has binomial
distribution,
whose expected value is $\expect{M}=M_s(1-q)$.
Let the matrix $\B{S}\in \mathcal{R}^{M \times N}$ contain the columns of the signatures from all neighbors of node $Z_0$, observable
during the $M$ off-slots of node $Z_0$. To ensure the expected value of the $l_2$ norm of each column in $\B{S}$ to be 1, we normalize
the signature matrix $\B{S}$ by $\sqrt{M_s (1-q)q}$.

We focus on the first frame received by $Z_0$ as the second frame is of
identical form.
The received signal of $Z_0$ during its $M$ off-slots based on~\eqref{eq:y0}
can be expressed as
\begin{align}\label{eq:Y0}
\B{Y}_0 = \sqrt{\gamma_s} \B{S} \B{X} + {\B{W}}_0,
\end{align}
where
\begin{align}
  \label{eq:5}
\gamma_s=\gamma M_s (1-q)q/\sigma^2,
\end{align}
$\B{W}_0$ consists of circularly symmetric complex Gaussian entries
with unit variance,
and $\B{X}$ is a binary $N$-vector for indicating which $K$ signatures are selected to form the
sum in \eqref{eq:y0} as well as the signal strength of each
neighbor. Specifically,
\begin{align}  \label{eq:3}
  \B{X}_{(i-1)2^l+j}=  U_{0i} \mathbf{1}(\omega_i=j)
\end{align}
for $i=1,\dots,K$ and $j=1,\dots,2^l$.
For example, consider $K=3$ neighbors with $l=2$ bits of information
each, where the messages are $\omega_1=1$, $\omega_2=3$ and
$\omega_3=2$,
the vector $\B{X}$ can be expressed as
\begin{align}  \label{eq:2}
  \B{X}=[U_{01}\; 0\;\;0\;\;0\quad0\;\;0\;\;U_{02}\;\;0\quad0\;\;U_{03}\;\;0\;\;0]^T.
\end{align}
The sparsity of the signal is exactly $2^{-l}$.



\subsection{The Effect of Interference}

As aforementioned, the impact of interference from non-neighbors is
accounted for in the variance ($\sigma^2$) of the additive noise term
in channel model~\eqref{eq:y0}.
The variance $\sigma^2$, or rather the
signal-to-interference-plus-noise ratio $\gamma_s$ in model~\eqref{eq:Y0},
will be used by the decoder.
Once the coding and transmission schemes have been described in
Section~\ref{s:enc}, $\sigma^2$ can be obtained as follows.

We first derive the aggregate interference caused by
non-neighbors of node $Z_0$ in each time slots as
\begin{align}\label{9}
&\expsub{\hat{\Phi}_q} { \mathop{\sum}_{\left(Z_i, \mathcal{G}_i\right)\in \hat{\Phi}_q} \gamma |h_{0i}|^2 R^{-\alpha}_{0i}
\mathbf{1}\left( |h_{0i}|^2 R^{-\alpha}_{0i} < \theta \right) } \nonumber
\\&\quad=2\pi\lambda q \gamma \int^\infty_0 \int^\infty_0 \bar{h} r^{-\alpha} \mathbf{1}\left( \bar{h} r^{-\alpha} <
\theta \right) r e^{-\bar{h}} \diff r \diff \bar{h} \nonumber
\\&\quad= 2\pi\lambda q\gamma \int^\infty_0 r^{-\alpha+1}\left[1-(\theta r^\alpha+1) e^{-\theta r^\alpha}\right]  \diff r \nonumber \\&\quad=\frac{4}{\alpha(\alpha-2)}\pi\lambda q \gamma \theta^{1-\frac{2}{\alpha}}\Gamma\left(\frac{2}{\alpha}\right),
\end{align}
where $\hat{\Phi}_q$ is an independent thinning of $\hat{\Phi}$ with
retention probability $q$, constructed  by those transmitting
nodes in each slot, which is still an independent marked Poisson point
process, but with intensity $\lambda q$. Therefore, the
variance is 
\begin{align}\label{10}
\sigma^2 = \frac{4}{\alpha(\alpha-2)}\pi \lambda q \gamma \theta^{1-\frac{2}{\alpha}} \Gamma\left(\frac{2}{\alpha}\right) + 1.
\end{align}

\subsection{Decoding via Sparse Recovery}
\label{s:dec}

\begin{algorithm} 
\caption{The Message-Passing Decoding Algorithm}
\label{algo:bp}
\begin{algorithmic}[1]
\STATE {\it Input:} $\B{S},\B{Y},\gamma_s,q$.
\STATE {\it Initialization:}
    \STATE $\Lambda \leftarrow -\log(2^l-1)$,
 $L \leftarrow \sum_{\mu}|\partial \mu|$, $L_2 \leftarrow \sum_{\mu}|\partial \mu|^2$
    \STATE $\B{R} \leftarrow 2\gamma_s^{-1/2}\B{Y}-\B{S} \cdot \B{1}$
    \STATE $\hat{m}_{\mu k}^0 \leftarrow 0$ for all $\mu,k$
\STATE {\it Main iterations:}
\FOR{$t=1$ to $T-1$}
    \FORALL{$\mu$, $k$ with $s_{\mu k} \neq 0$}
        \STATE $m_{k\mu}^t \leftarrow \tanh\big(\frac{\Lambda}{2}+\sum_{\nu \in \partial k \backslash \mu}\tanh^{-1}\hat{m}_{\nu k}^{t-1} \big)$
    \ENDFOR
    \STATE $Q^t \leftarrow \frac{1}{L_2}\sum_{\mu} |\partial \mu|\sum_{j \in \partial \mu}(m_{j\mu}^t)^2$
    \STATE $A^t \leftarrow \big[\frac{4}{\gamma_s}+\frac{L_2}{M_s(1-q)qL}(1-Q^t)\big]^{-1}$
    \FORALL{$\mu$, $k$ with $s_{\mu k} \neq 0$}
        \STATE $\hat{m}_{\mu k}^t \leftarrow \tanh \big(A^ts_{\mu k}( r_{\mu}-\sum_{j \in \partial \mu \backslash k}s_{\mu j}m_{j\mu}^t)\big)$
    \ENDFOR
\ENDFOR
\STATE $m_k \leftarrow \tanh\big(\frac{\Lambda}{2}+\sum_{\nu \in
  \partial k}\tanh^{-1}\hat{m}_{\nu k}^{T-1} \big)$ for all $k$
\STATE {\it Output:} for  $i=1,\dots,K$,\\
${w}_i = \mathop{\arg\max}\limits_{j\in\{1,\dots,2^l\}} |m_{(i-1)2^l+j}|$\\
${u}_i = \mathop{\max}\limits_{j\in\{1,\dots,2^l\}} |m_{(i-1)2^l+j}|$
\end{algorithmic}
\end{algorithm}

In this subsection, we discuss how to process the received signal
$\B{Y}_0$ described by~\eqref{eq:Y0} to recover signal $\B{X}$, which
contains location messages from all
neighbors as well as the corresponding amplitudes that indicate their
distances.  The signature matrix $\B{S}$, the
signal-to-interference-plus-noise ratio $\gamma_s$, as well as the
duty cycle of the signatures $q$ are all inputs to the decoder.

To recover the sparse signal $\B{X}$ based on observations made
through linear model~\eqref{eq:Y0} is fundamentally a problem of
compressed sensing.
There have been a number of algorithms~\cite{needell2009applied,
donoho2010message} developed in the literature to solve the problem,
the complexity of which are often polynomial in the size of the
codebook. In this paper, we augment the iterative algorithm proposed in
previous work~\cite{zhang2013virtual}.  The difference is that
in~\cite{zhang2013virtual}, the decoder only outputs the support of
$\B{X}$, whereas in this work, we also estimate the amplitudes of the
non-zero elements of $\B{X}$.
The iterative message-passing algorithm is based on belief
propagation.  The computational complexity is in the order of
$\mathcal{O}(M Nq)$, the same order as the complexity of other two other
popular sparse recovery algorithms~\cite{needell2009applied,
  donoho2010message}.

The message-passing algorithm, described as Algorithm~\ref{algo:bp},
solves an inference problem on a Forney-style
bipartite factor graph that represents the model~\eqref{eq:Y0}.
Since~\eqref{eq:Y0} is complex valued, we divide it into the real and
imaginary parts,
which share the same bipartite graph.
Let us focus on one of them:
\begin{align}\label{12}
  y_{\mu} = \sqrt{\gamma_s} \mathop{\sum}^N_{k=1} s_{\mu k} x_k + w_\mu,
\end{align}
where $\mu=1,...,M$ and $k=1,...,N$ index the measurements (the off
slots within one frame) and the ``input symbols,'' respectively.
$x_k$ (resp. $y_\mu$ ) corresponds to the value of the symbol node
(resp. measurement node). For every $(\mu,k)$, there is a link
between symbol node $k$ and measurement node $\mu$ if $s_{\mu k}\neq
0$. For convenience, let $\partial{\mu}$ (resp. $\partial{k}$) denote
the subset of symbol nodes (resp. measurement nodes) connected
directly to measurements node $\mu$ (resp. symbol node $k$), called
its neighborhood.\footnote{This is to be distinguished from the notion
  of neighborhood in the wireless network defined in Section II-B.}
Also, let $\partial{\mu} \backslash k$ denote the
neighborhood of measurement node $\mu$ excluding symbol node $k$ and
let $\partial{k} \backslash \mu$ be similarly defined.

In each iteration of Algorithm~\ref{algo:bp}, every symbol node $k$
first computes messages to pass to the measurement nodes connected to
them; then every measurement node $\mu$ computs messages to pass to
their corresponding symbol nodes.  After $T$ iterations (typically about
10 iterations suffice), Algorithm~\ref{algo:bp} outputs two
sets of estimates.
One set consists of the position of the largest element of each of $K$
sub-vectors, corresponding to the location message ($\omega_i$) from
neighbor $Z_i$;
the other set consists of the corresponding amplitude ($|u_i|$) of the largest
element of each of $K$
sub-vectors, which shall be used to infer about the distance to
neighbor ($Z_i$).


\subsection{Distance Estimation}
\label{s:distance}


Each node estimates its distances to neighbors using the signal
strength estimates produced by the decoder.
Again, we focus on node $Z_0$.
For simplicity, we assume that node $Z_0$ knows or can estimate
(through training) the fading gain
$|h_{0i}|^2$ between itself and each neighbor $Z_i$.
Based on the relation between the distance and the received signal
amplitude given by~\eqref{eq:U0i}, the distance between
neighboring nodes $Z_0$ and $Z_i$ can be estimated as
\begin{align}  \label{eq:d0i}
  r_{0i}=\left(\frac{u_{0i}^2}{|h_{0i}|^2}\right)^{-1/\alpha},
\end{align}
where $u_{0i}$ is the estimated channel coefficient ($|u_{0i}|$the received signal
amplitude) of $U_{0i}$ in~\eqref{eq:U0i}.


\subsection{Location Estimation via Convex Optimization}


Using the procedures described in Sections~\ref{s:dec}
and~\ref{s:distance}, each node decodes the locations of all its
neighbors (or their current estimates),
as well as the
approximate distances to those neighbors.
The node then estimates its own location as the point on the plane
that is the most consistent with all information the node has
collected.

Let us again focus on node $Z_0$.
The node has acquired $(z_i, u_{i0})$ for all neighbors $Z_i\in\mathcal{N}(Z_0)$.
We let the location estimate of $Z_0$ be $z_0\in\reals^2$ which minimizes the
following error:
\begin{align}\label{N-1}
  \sum_{Z_i\in \mathcal{N}(Z_0)}
  \left| \|z_0-z_i\|^2
  -
  r^2_{0i} \right|.
\end{align}
If all estimates were perfect, then $r_{0i} = \|z_0-z_i\|$, so that
the error is equal to 0.
In general, the estimates are imperfect due to noise and interference,
as well as possibly lack of network connectivity.

The preceding minimization problem is non-convex.
Following~\cite{srirangarajan2008distributed},
we use relaxation to turn the problem into a second order cone programming
(SOCP) problem~\cite{boyd2004convex}.
Relaxing the equality constraints in  to ``greater than or equal to'' inequality constraints,
we obtain the following convex problem in an SOCP form as
\begin{align}\label{eq:socp}
  \mathop{\text{minimize}}\limits_{z_0, y_{0i}, t_{0i}}    \;\;
  & \mathop{\sum}_{Z_i \in \mathcal{N}(Z_0)} t_{0i} \nonumber
  \\
  \text{subject to} \;\;
  & y_{0i} \geq \|z_0-z_i\|^2 \text{ and }
  t_{0i} \geq |y_{0i}-r^2_{0i}| \nonumber \\
  & \text{ for all } Z_i \in \mathcal{N}(Z_0).
\end{align}
This optimization problem can be solved efficiently to yield an
estimate ($z_0$) of the location of node $Z_0$.

\subsection{Summary of the Algorithm}

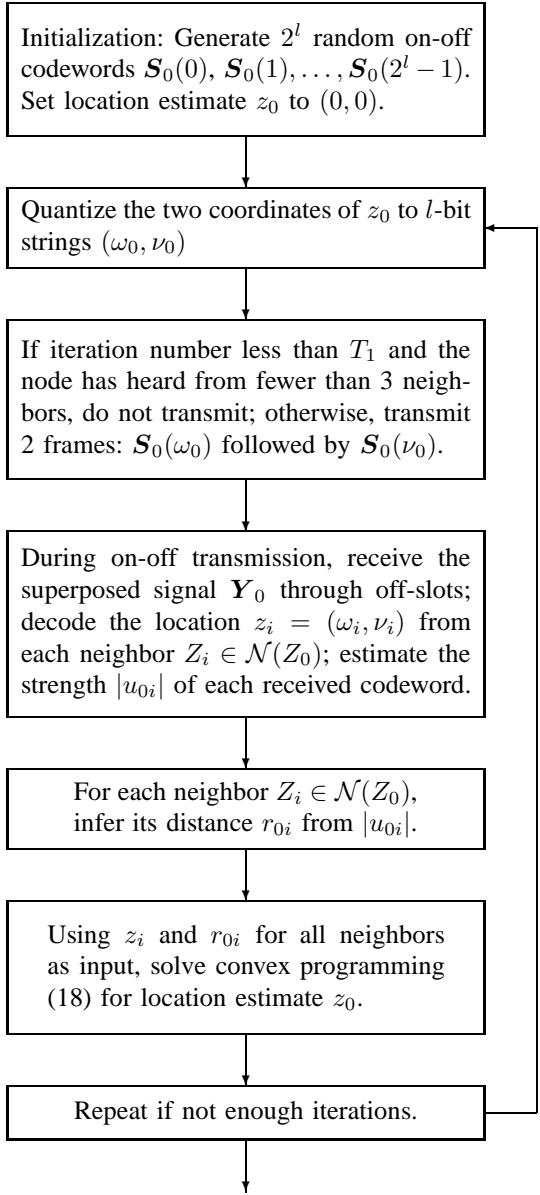
\begin{figure}
  \centering
  \begin{picture}(190,455)
     \put(0,400){\framebox(180,50){\parbox{170pt}{Initialization:
           Generate $2^l$
           random on-off codewords
           $\B{S}_0(0)$, $\B{S}_0(1),\dots,\B{S}_0(2^l-1)$.
           Set location estimate $z_0$ to $(0, 0)$.}}}
    \put(90,400){\vector(0,-1){20}}
    \put(0,350){\framebox(180,30){\parbox{170pt}{Quantize the two coordinates of $z_0$ to $l$-bit strings $\left(\omega_0, \nu_0\right)$}}}
    \put(90,350){\vector(0,-1){20}}
    \put(0,270){\framebox(180,60){\parbox{170pt}{If iteration number
          less than $T_1$ and the node has heard from fewer than 3
          neighbors, do not transmit; otherwise, transmit 2 frames:
          $\B{S}_0(\omega_0)$ followed by $\B{S}_0(\nu_0)$. 
        }}}
    \put(90,270){\vector(0,-1){20}}
    \put(0,180){\framebox(180,70){\parbox{170pt}{During on-off
          transmission, receive the superposed signal $\B{Y}_0$ through off-slots;
          decode the location $z_i=(\omega_i, \nu_i)$
          from each neighbor $Z_i\in \mathcal{N}(Z_0)$; estimate the
          strength $|u_{0i}|$ of each received codeword.
}}}
    \put(90,180){\vector(0,-1){20}}
    \put(0,130){\framebox(180,30){\parbox{130pt}{For each neighbor
          $Z_i\in\mathcal{N}(Z_0)$, infer its distance $r_{0i}$ from $|u_{0i}|$.}}}
    \put(90,130){\vector(0,-1){20}}
    \put(0,60){\framebox(180,50){\parbox{150pt}{Using $z_i$ and
          $r_{0i}$ for all neighbors as input, solve convex
          programming \eqref{eq:socp} for location estimate $z_0$.}}}
    \put(90,60){\vector(0,-1){20}}
    \put(0,20){\framebox(180,20){Repeat if not enough iterations.}}
   \put(90,20){\vector(0,-1){20}}
    \put(180,30){\line(1,0){20}}
    \put(200,30){\line(0,1){335}}
    \put(200,365){\vector(-1,0){20}}
  \end{picture}
  \caption{The flow chart of the algorithm carried out by node $Z_0$.}
  \label{f:chart}
\end{figure}

In the flow chart depicted in Fig.~\ref{f:chart}, we summarize the
procedure every node executes in distributed manner to accomplish
network-wide localization.
At the beginning, all clients assume they are located at the origin $(0,0)$.
Every node generates a random codebook with $2^l$ codewords, each of
$M_s$ symbols.  The codewords of node $Z_i$ are
$\B{S}_i(0),\dots,\B{S}_i(2^l-1)$.

Localization is carried out iteratively in two stages.
In each iteration, all anchors transmit their quantized location coordinates.
In the first stage (i.e., in the first $T_1$ iterations), a client
transmits its quantized coordinates if it has heard from three or more
neighbors in the previous iteration.
In the second stage (after $T_1$ iterations), all clients
transmit their quantized coordinates
regardless of how many neighbors they have heard from.
We restrain some clients from transmission in the first stage because
their location estimates are poor since they have heard from only two
or fewer neighbors.
At the end of each iteration, all clients can estimate its own location based on the received signal.

Without loss of generality, we focus on node $Z_0$ next and describe its
actions.
In each iteration, if node $Z_0$ transmits, it quantizes its two coordinates to
$(\omega_0, \nu_0)$, each using $l$ bits.  It then transmits two
codewords $\B{S}_0(\omega_0)$ and  $\B{S}_0(\nu_0)$ in two consecutive
frames.
A client receives the superposed signals from its neighbors through
its own off-slots.
The node decodes the location message $(\omega_i,\nu_i)$ from
neighbor $Z_i$ and also infers the distance to $Z_i$ based
on the estimated signal amplitude $|{u}_{0i}|$. 
Based on the locations of and distances to neighbors, the client $Z_0$
estimates its location $z_0$ by solving the optimization
problem~\eqref{eq:socp}.
The preceding procedure is repeated by every node until convergence or
until a given number of iterations is finished.


\section{Numerical Results}
\label{s:num}

In this section, we evaluate the performance of the proposed
localization algorithm through simulations.
We use the CVX MATLAB toolbox 
for solving the optimization problem~\eqref{eq:socp}.

Without loss of generality, let one unit of distance be 1 meter. Consider a wireless network of 100 nodes. We randomly
generate the true locations of client nodes according to a uniform distribution on the square $[0,50]\times[0,50]$.
The nodes form a Poisson point process in the square conditioned on the node
population. Suppose the path-loss exponent $\alpha=3$. The threshold of channel gain to define neighborhood is set to
$\theta = 10^{-3}$.  It means that if the transmit power for a node one meter away is 30 dB, then the SNR attenuates to 0 dB
at $= 10^{3/\alpha}=10$ meters in the absence of fading, i.e., the
coverage of the neighborhood of a node is typically a circle of
radius 10 meters. According to \eqref{6}, a node near the center of the square (without boundary effect) has on average $c\approx 11$ neighbors.
Here we use the nominal SNR in the following simulations, indeed the SNR will decrease at most 30 dB at the revived node due to the path loss in a neighborhood circle.
Each node quantizes its location to 8 bits per coordinate, i.e., about
0.4 meters in precision.
The node then encodes
each quantized coordinate to an on-off codeword of 600 symbols.
Each frame is thus of 600 symbol intervals, and each iteration
is of 1,200 symbol intervals.

\begin{figure}\centering
    \includegraphics[width=\columnwidth]{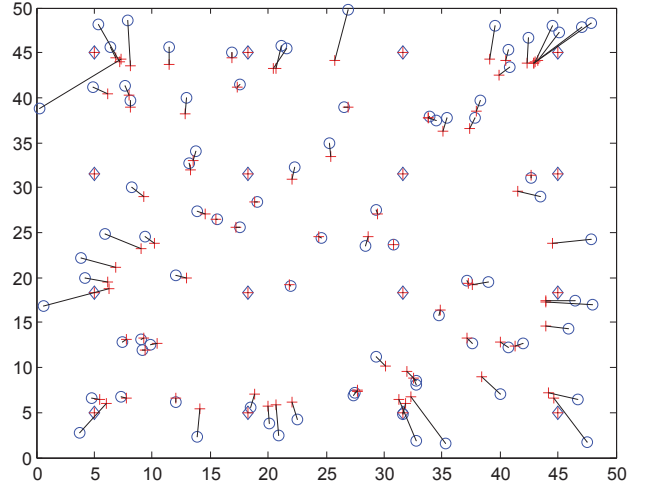}
    \caption{Location results for 100 nodes on $[0 ~ 50]^2$, including
      16 anchors forming a 4$\times$4 lattice,  SNR=30 dB.}
    \label{fixed_topology}
\end{figure}

\begin{figure}\centering
    \includegraphics[width=\columnwidth]{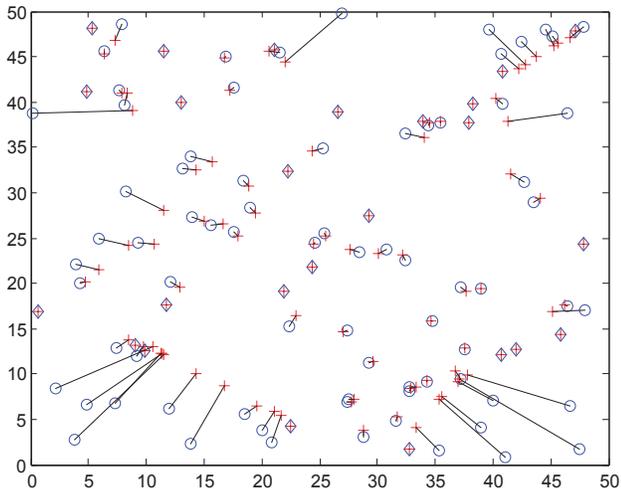}
    \caption{Locations results for uniform topology: $[0 \quad 50]^2$, 100 nodes, including 25 randomly generated anchors, SNR=30dB  }
    \label{random_topology}
\end{figure}

Figs.~\ref{fixed_topology} and~\ref{random_topology} demonstrate the
location results for a network of 100 nodes in two scenarios.
The true locations of clients and anchors are depicted using $\circ$ and
$\diamond$, respectively. The estimated node locations are depicted
using $+$. Solid lines indicate the error
between the true locations and the estimated locations.
In Fig.~\ref{fixed_topology}, the network consists of 16 anchors
forming a $4\times4$ lattice, where the remaining 84 clients are Poisson
distributed.
In Fig.~\ref{random_topology}, the network consists of 25 anchors and
75 clients, all Poisson distributed.

 In Fig.~\ref{fixed_topology}, with the assistance 16 anchors forming a
lattice, all
clients in the convex hull of their neighbors can accurately estimate
their own locations.  The error is typically a small fraction of a meter.
The estimated locations are less accurate for clients near the
boundaries because they generally have fewer neighbors.  This is thus
a boundary effect.

In Fig.~\ref{random_topology}, with the assistance of 25 randomly
placed anchors, the accuracy of the location estimates
is about the same as in Fig.~\ref{fixed_topology} for nodes near the
center of the network.
The boundary effect is slightly more pronounced in
Fig.~\ref{random_topology} than in Fig.~\ref{fixed_topology}, although
more anchors are adopted.  This is mainly because fewer anchors are
found near the lower left and lower right corners.

\begin{figure}\centering
    \includegraphics[width=\columnwidth]{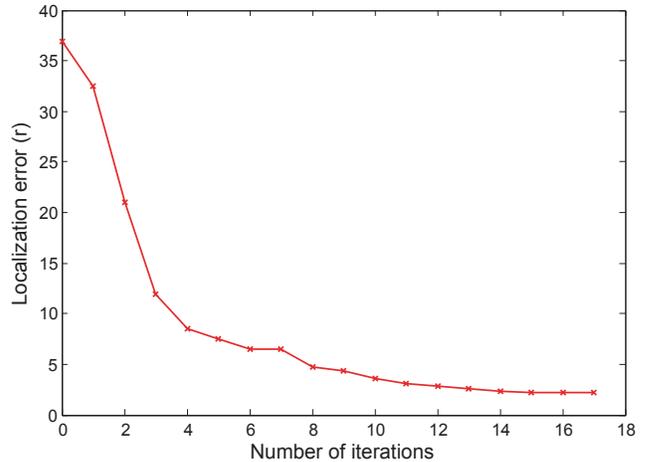}
    \vspace*{-3mm}
    \caption{Average localization error vs iterations, 100 nodes, including
      16 anchors forming a 4$\times$4 lattice, SNR=30dB }
    \label{iteration}
    \vspace*{-1mm}
\end{figure}

\begin{figure}\centering
    \includegraphics[width=\columnwidth]{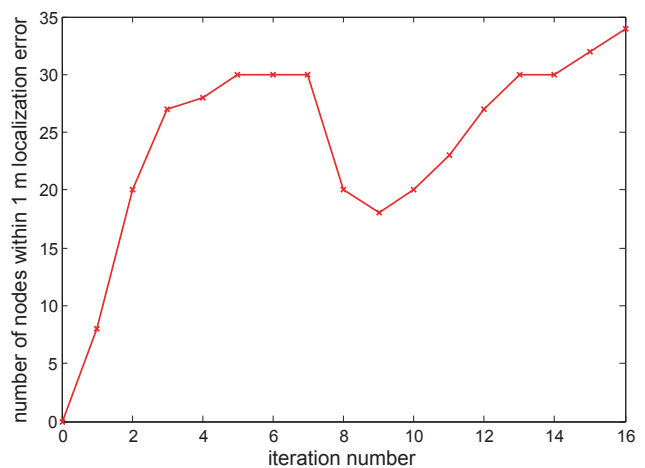}
    \caption{Number of nodes within 1 meter localization error vs iterations, 100 nodes, including
      16 anchors forming a 4$\times$4 lattice, SNR=30dB }
    \label{num_1m_ite}
\end{figure}

Let the average localization error of all clients be calculated as
\begin{align}  \label{eq:1}
\frac1{|\Phi_c|} \sum^{|\Phi_c|}_{i=1} \|Z_i-z_i\|
\end{align}
where $Z_i$ and $z_i$ denote the true location and the estimated
location of node $Z_i$, respectively.  Figs.~\ref{iteration}
and~\ref{num_1m_ite} are based on the same network realization.  In
Fig.~\ref{iteration}, the average localization error decreases
monotonically with the number of iterations.
In the first stage, i.e., iterations 1 through 7, we only let clients
who have heard from three or more neighbors transmit.
The location error improves quickly in the first stage.
Here we determine the number of iterations in the first stage in a
global matter (in practice this should be determined {\em a priori}):
If no new client join the set of nodes who have heard three or more
neighbors in an iteration, we shall move to the second stage.

As shown in Fig.~\ref{num_1m_ite}, the number of clients within 1
meter localization error rises quickly in the first stage.  Starting
from iteration 8, the number decreases for a few iterations until it
rises again.  The reason for this is as follows: At the beginning of
stage 2, as we let clients who have only heard from two or fewer
neighbors in the previous iteration transmit their location estimates,
most clients' estimate improves in the expense of some nodes who
already have very accurate estimates.
With additional iterations, all clients eventually converge to good
estimates of their locations as shown in Fig.~\ref{iteration} and
Fig.~\ref{num_1m_ite}.

\begin{figure}\centering
    \includegraphics[width=\columnwidth]{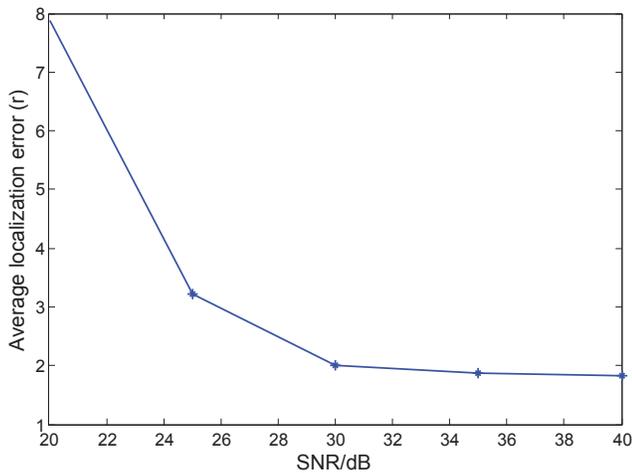}
    \caption{Average localization error vs SNR, 100 nodes, including
      16 anchors forming a 4$\times$4 lattice. }
    \label{error_SNR}
\end{figure}

Fig.~\ref{error_SNR} shows the average localization error of the
proposed algorithm versus SNR.  
The average localization error decreases sharply with the SNR in the
noise-dominant regime (SNR below 30 dB).  As the SNR rises above 30
dB, interference dominates, so that the performance barely improves as
the SNR further increases.



\section{Concluding Remarks}
\label{s:con}

In this paper, we have proposed a complete, distributed, iterative
solution for ranging and localization in wireless networks.
Importantly, it is recognized that local-area functions such as
ranging and localization are best addressed in the physical layer to
exploit the broadcast and multiaccess nature of the wireless medium.
The proposed coding and estimation techniques based on compressed sensing
have shown to be highly effective through simulations.
In particular, each iteration requires 1,200 symbol transmissions and
about 10 iterations suffice.  Note that a single WiFi frame typically
consists of several thousand symbols.
Thus the proposed scheme is considerably more efficient than
conventional schemes based on random access, where many more
transmissions and retransmissions are needed due to collisions.
It would of course be interesting to evaluate the algorithm under more
realistic scenarios, e.g., using a software radio implementation.
This is left to future work.


\bibliographystyle{IEEEtran}
\bibliography{IEEEabrv,gan2013distributed}

\begin{thebibliography}{10}
\providecommand{\url}[1]{#1}
\csname url@samestyle\endcsname
\providecommand{\newblock}{\relax}
\providecommand{\bibinfo}[2]{#2}
\providecommand{\BIBentrySTDinterwordspacing}{\spaceskip=0pt\relax}
\providecommand{\BIBentryALTinterwordstretchfactor}{4}
\providecommand{\BIBentryALTinterwordspacing}{\spaceskip=\fontdimen2\font plus
\BIBentryALTinterwordstretchfactor\fontdimen3\font minus
  \fontdimen4\font\relax}
\providecommand{\BIBforeignlanguage}[2]{{%
\expandafter\ifx\csname l@#1\endcsname\relax
\typeout{** WARNING: IEEEtran.bst: No hyphenation pattern has been}%
\typeout{** loaded for the language `#1'. Using the pattern for}%
\typeout{** the default language instead.}%
\else
\language=\csname l@#1\endcsname
\fi
#2}}
\providecommand{\BIBdecl}{\relax}
\BIBdecl

\bibitem{steiniger2006foundations}
M.~N. S.~Steiniger and A.~Edwardes, ``Foundations of location based services,''
  in \emph{Lecture Notes on LBS 1}, 2006.

\bibitem{mohapatra2008survey}
D.~Mohapatra and S.~B. Suma, ``Survey of location based wireless services,'' in
  \emph{Proc.\ International Conference on Personal Wireless
  Communications}.\hskip 1em plus 0.5em minus 0.4em\relax Sydney, Australia,
  2008, pp. 189--194.

\bibitem{surie2005wireless}
D.~Surie, O.~Laguionie, and T.~Pederson, ``Wireless sensor networking of
  everyday objects in a smart home environment,'' in \emph{Proc.\ International
  Conference on Intelligent Sensors, Sensor Networks and Information
  Processing}, 2005, pp. 358--362.

\bibitem{kim2009performance}
H.~Kim, ``Performance comparison of asynchronous ranging algorithms,'' in
  \emph{Proc.\ IEEE GLOBECOM}.\hskip 1em plus 0.5em minus 0.4em\relax Hawaii,
  USA, 2009.

\bibitem{hwang2011AVSR}
K.~Hwang, ``Avsr: Asynchronous virtual slot-based ranging for user-oriented
  location services,'' \emph{IEEE Transactions on Consumer Electronics},
  vol.~57, no.~1, pp. 203--208, February 2011.

\bibitem{fu2007ranging}
X.~Fu, Y.~Li, and H.~Minn, ``A new ranging method for ofdma systems,''
  \emph{IEEE Transactions on Wireless Communications}, vol.~6, no.~2, pp.
  659--669, February 2007.

\bibitem{srirangarajan2008distributed}
S.~Srirangarajan, A.~H. Tewfik, and Z.~Luo, ``Distributed sensor network
  localization using ocp relaxation,'' \emph{IEEE Transactions on Wireless
  Communications}, vol.~7, no.~12, pp. 4886--4895, December 2008.

\bibitem{qingjiang2010distributed}
Q.~Shi, C.~He, H.~Chen, and L.~Jiang, ``Distributed wireless sensor network
  localization via sequential greedy optimization algorithm,'' \emph{IEEE
  Transactions on Signal Processing}, vol.~58, no.~6, pp. 3328--3340, June
  2010.

\bibitem{chiu2012robust}
W.~Chiu, B.~Chen, and C.~Yang, ``Robust relative location estimation in
  wireless sensor networks with inexact position problems,'' \emph{IEEE
  Transactions on Mobile Computing}, vol.~11, no.~6, pp. 935--946, June 2012.

\bibitem{zhang2013virtual}
L.~Zhang and D.~Guo, ``Virtual full duplex wireless broadcast via sparse
  recovery,'' \emph{submitted to IEEE Transactions Networking, revised}, 2013,
  http://arxiv.org/abs/1101.0294.

\bibitem{donoho2006compressed}
D.~L. Donoho, ``Compressed sensing,'' \emph{IEEE Transactions on Information
  Theory}, vol.~52, no.~4, pp. 1289--1306, April 2006.

\bibitem{candes2008introduction}
E.~J. Candes and M.~B. Wakin, ``An introduction to compressive sampling,''
  \emph{IEEE Signal Processing Magazine}, vol.~25, no.~2, pp. 21--30, March
  2008.

\bibitem{candes2005decoding}
E.~J. Candes and T.~Tao, ``Decoding by linear programming,'' \emph{IEEE
  Transactions on Information Theory}, vol.~51, no.~12, pp. 4203--4215, Dec.
  2005.

\bibitem{guo2010virtual}
D.~Guo and L.~Zhang, ``Virtual full-duplex wireless communication via rapid
  on-off-division duplex,'' in \emph{Proc.\ Allerton Conf. Commun., Control, \&
  Computing, Monticello}.\hskip 1em plus 0.5em minus 0.4em\relax IL, USA, 2010.

\bibitem{zhang2013neighbor}
L.~Zhang, J.~Luo, and D.~Guo, ``Neighbor discovery for wireless networks via
  compressed sensing,'' \emph{Performance Evaluation}, vol.~70, pp. 457--471,
  2013.

\bibitem{schizas2008consensus}
I.~D. Schizas, A.~Ribeiro, and G.~B. Giannakis, ``Consensus in ad hoc wsns with
  noisy links-part i: Distributed estimation of deterministic signals,''
  \emph{IEEE Trans. Signal Process.}, vol.~56, no.~1, pp. 350--364, 2008.

\bibitem{simeone2008distributed}
O.~Simeone, U.~Spagnolini, Y.~Bar-Ness, , and S.~Strogatz, ``Distributed
  synchronization in wireless networks,'' \emph{IEEE Signal Processing Mag.},
  vol.~25, no.~5, pp. 81--97, 2008.

\bibitem{baccelli2009stochastic}
F.~Baccelli and B.~Blaszczyszyn, ``Stochastic geometry and wireless networks:
  Volume i theory and volume ii applications,'' in \emph{vol.4 of Foundations
  and Trends in Networking}.\hskip 1em plus 0.5em minus 0.4em\relax NoW
  Publishers, 2009, vol.~4.

\bibitem{borbash2007asynchronous}
S.~A. Borbash, A.~Ephremides, and M.~J. McGlynn, ``An asynchronous neighbor
  discovery algorithm for wireless sensor networks,'' \emph{Ad Hoc Networks},
  vol.~5, no.~7, pp. 998--1016, Dec. 2007.

\bibitem{needell2009applied}
D.~Needell and J.~A. Tropp, ``Cosamp: Iterative signal recovery from incomplete
  and inaccurate samples,'' \emph{Applied and Computational Harmonic Analysis},
  vol.~26, no.~3, pp. 301--321, May 2009.

\bibitem{donoho2010message}
D.~L. Donoho, A.~Maleki, and A.~Montanari, ``Message passing algorithms for
  compressed sensing: I. motivation and construction and ii,'' in \emph{Proc.\
  IEEE Inform. Theory Workshop}.\hskip 1em plus 0.5em minus 0.4em\relax Cairo,
  Egypt, 2010.

\bibitem{boyd2004convex}
S.~Boyd and L.~Vandenberghe, \emph{Convex Optimization}.\hskip 1em plus 0.5em
  minus 0.4em\relax Cambridge University Press, 2004.

\end{thebibliography}

\end{document}